\documentclass[10pt,letterpaper]{article}
\usepackage[top=0.85in,left=2.75in,footskip=0.75in]{geometry}

\usepackage{amsmath,amssymb}

\usepackage{changepage}

\usepackage[utf8x]{inputenc}

\usepackage{textcomp,marvosym}

\usepackage{cite}

\usepackage[right]{lineno}

\usepackage{microtype}
\DisableLigatures[f]{encoding = *, family = * }

\usepackage[table]{xcolor}

\usepackage{array}

\newcolumntype{+}{!{\vrule width 2pt}}

\usepackage{mdframed}

\newlength\savedwidth

\raggedright
\usepackage[aboveskip=1pt,labelfont=bf,labelsep=period,justification=raggedright,singlelinecheck=off]{caption}

\makeatletter
\renewcommand{\@biblabel}[1]{\quad#1.}
\makeatother

\date{}

\usepackage{lastpage,fancyhdr,graphicx}
\usepackage{epstopdf}
\fancyheadoffset[L]{2.25in}
\fancyfootoffset[L]{2.25in}
\newcommand{\rulemajor}[2]{\section{#1}\label{#2}}
\newcommand{\ruleref}[1]{Rule~\ref{#1}}

\begin{document}
\vspace*{0.2in}

% Title must be 250 characters or less.
\begin{flushleft}
{\Large
\textbf\newline{Ten simple rules for collaborative lesson development} % Please use "sentence case" for title and headings (capitalize only the first word in a title (or heading), the first word in a subtitle (or subheading), and any proper nouns).
}
\newline
% Insert author names, affiliations and corresponding author email (do not include titles, positions, or degrees).
\\
{Gabriel~A.~Devenyi}\textsuperscript{1{\ddag}},
{R\'{e}mi~Emonet}\textsuperscript{2{\ddag}},
{Rayna~M.~Harris}\textsuperscript{3{\ddag}},
{Kate~L.~Hertweck}\textsuperscript{4{\ddag}},
{Damien~Irving}\textsuperscript{5{\ddag}},
{Ian~Milligan}\textsuperscript{6{\ddag}},
{Greg~Wilson}\textsuperscript{7{\ddag}*}
\\
\bigskip
\textbf{1} Douglas Mental Health University Institute, McGill University / gdevenyi@gmail.com \\
\textbf{2} Univ Lyon, UJM-Saint-Etienne, F-42023, France / remi.emonet@univ-st-etienne.fr \\
\textbf{3} The University of Texas at Austin / rayna.harris@utexas.edu \\
\textbf{4} The University of Texas at Tyler / khertweck@uttyler.edu \\
\textbf{5} CSIRO Oceans and Atmosphere / irving.damien@gmail.com \\
\textbf{6} University of Waterloo / i2millig@uwaterloo.ca \\
\textbf{7} Rangle.io / gvwilson@third-bit.com \\

\bigskip

% Insert additional author notes using the symbols described below. Insert symbol callouts after author names as necessary.
%
% Remove or comment out the author notes below if they aren't used.
%
% Primary Equal Contribution Note
{\ddag} These authors contributed equally to this work.

% Additional Equal Contribution Note
% Also use this double-dagger symbol for special authorship notes, such as senior authorship.
% \ddag These authors also contributed equally to this work.

% Current address notes
% \textcurrency Current Address: Dept/Program/Center, Institution Name, City, State, Country % change symbol to "\textcurrency a" if more than one current address note
% \textcurrency b Insert second current address
% \textcurrency c Insert third current address

% Deceased author note
% \dag Deceased

% Group/Consortium Author Note
% \textpilcrow Membership list can be found in the Acknowledgments section.

% Use the asterisk to denote corresponding authorship and provide email address in note below.
* corresponding author

\end{flushleft}

% Please keep the abstract below 300 words

% \linenumbers

\section*{Abstract}

The collaborative development methods pioneered by the open source software community
offer a way to create lessons that are open, accessible, and sustainable.
This paper presents ten simple rules for doing this
drawn from our experience with several successful projects.

% Please keep the Author Summary between 150 and 200 words
% Use first person. PLOS ONE authors please skip this step.
% Author Summary not valid for PLOS ONE submissions.
\section*{Author summary}

Lessons take significant effort to build and  maintain.
We have found that
the collaborative development methods pioneered by the open source software community
are an effective, economical way to create and sustain lessons
that can be used by large numbers of people in a wide variety of contexts.
The ten simple rules outlined in this paper
summarize the best practices that we have helped evolve
in several successful open education projects aimed at researchers and research librarians
in a wide range of disciplines.

% Use "Eq" instead of "Equation" for equation citations.

\section*{Introduction}

Lessons take significant effort to build and even more to maintain.
Most academics do this work on their own,
but leveraging a community approach
can make educational resource development more sustainable, robust, and responsive.
Treating lessons as a community resource
to be updated, adapted, and improved incrementally,
can free up valuable time while increasing quality.

Despite the success of openness in software development and the curation of Wikipedia,
it is an uncommon approach in academic settings.
Each year,
thousands of university lecturers teach subjects ranging from first year biology,
to graduate-level courses in Indian film.
Some use textbooks written by one or a few authors,
but beyond that,
they develop and maintain their course materials in isolation.
This is curious given that research depends critically on sharing,
and that most researchers complain about how much time teaching takes away from research,
but the sociology and psychology behind this blind spot are out of the scope of this paper.

The authors have many years of experience with community-developed lessons
in the context of research computing in the sciences and humanities
through organizations like Software Carpentry and Programming Historian.
Software Carpentry was founded in 1998 to teach scientists basic computing skills,
and has since spawned two sibling organizations called Data Carpentry and Library Carpentry.
Programming Historian was founded in 2008,
and has evolved into a collaboratively-edited site providing lessons to humanities scholars.
Their guiding principles are that lessons should be:

\begin{enumerate}

\item
  open and easily accessible, and

\item
  continually maintained, refined, and improved
  by a community of contributors.

\end{enumerate}

All open education projects satisfy the first criterion by definition,
but few satisfy the second.
While their lessons are occasionally updated by a small team
(as happens when a new edition of a book is edited and published),
this is not the same as continuous improvement by a large community of contributors.
The ten simple rules that follow summarize what we have learned about doing that
as maintainers, editors, and reviewers of lessons used by tens of thousands of people (Fig. 1, 2).

\subsection*{Acknowledgments}

We are grateful to everyone who provided feedback on this paper,
including
James Baker,
Nathan Moore,
Pariksheet Nanda,
Tom Pollard,
Byron Smith,
and Andrew Walker.
We are also grateful to the hundreds of people who have contributed to
Programming Historian, Data Carpentry, Software Carpentry, and Library Carpentry
over many years.

\rulemajor{Clarify your audience}{audience}

The first requirement for building lessons together is
to know who they are being built for.
``Archaeology students'' is far too vague:
are you and your collaborators thinking of
first-year students who need an introduction to the field,
graduate students who intend to specialize in the sub-discipline which is the lesson's focus,
or someone in between?
If different contributors believe different things about prerequisite knowledge,
equipment or software required,
or how much time learners will have,
they will find it difficult to work together.

Instead of starting with learning objectives (\ruleref{practices}),
it can be helpful to write \emph{learner profiles} to clarify
the learner's general background,
what they already know,
what \emph{they} think they want to do,
how the material will help them,
and any special needs they might have.
This technique is borrowed from user interface design,
and a typical learner profile is presented in Box~1.

\rulemajor{Make lessons modular}{modular}

Every instructor's needs are different,
so build small chunks that can be re-purposed in many ways.
A university lecturer in meteorology,
for instance,
might construct a course for their students by bringing together lessons on differential equations,
fluid mechanics,
and absorption spectroscopy.
Creating courses this way shifts the instructor's burden from writing to finding and synthesizing,
which are easier if lessons clearly define what they cover (\ruleref{audience}),
and if lessons have been designed by people with a shared world-view (\ruleref{practices}).

One way to achieve this is to take existing courses and break them down into smaller, single-purpose modules
(a change which has pedagogical and administrative advantages in its own right).
When this is done,
these modules can be made more discoverable
by referencing specific points in the model curricula promulgated by many professional societies
(e.g., as learning objectives).
Smaller modules are also more approachable for new contributors (\ruleref{empower}).

\rulemajor{Teach best practices for lesson development}{practices}

Decades of pedagogical research has yielded many insights into
how best to build and deliver lessons \cite{hlw}.
Unfortunately,
many college and university faculty have little or no training in education \cite{brownell},
so this knowledge is rarely applied in the classroom.

Our experience is that even a brief introduction to a few key practices
helps collaborative lesson development.
If people have a shared understanding of how lessons should be developed,
it is easier for them to work together.
Less obviously,
if people have a shared model of how lessons will be \emph{used},
they are more likely to build reusable material.
Finally,
teaching people how to teach is a great way to introduce them to each other and build community.

One popular lesson development methodology is presented in \cite{wiggins-mctighe}.
When this is used,
lessons are built by:

\begin{enumerate}

  \item
    identifying learning objectives,

  \item
    creating \emph{summative assessments} to determine whether those objectives have been met,

  \item
    designing \emph{formative assessments} to gauge learners' progress
    and give them a chance to practice key skills,

  \item
    putting those formative assessments in order,

  \item
    and only then writing lessons to connect each to the next.

\end{enumerate}

\noindent
This method is effective in its own right,
but its greatest benefit is that it gives everyone a framework for collaboration.

An example of how to teach these practices is Software Carpentry's instructor training program.
First offered in 2012,
is now a two-day course delivered both in-person and online
\cite{lessons-learned,instructor-training,how-to-teach-programming}.
This program teaches good pedagogical practices,
and introduces everyone who takes it to who Software Carpentry's lessons are for,
how they are delivered,
and how they are maintained.
Largely as a result of this training,
several hundred people per year now contribute to Software Carpentry's lessons.

\rulemajor{Encourage and empower contributors}{empower}

Making the process for contributing to a lesson simple and transparent
is the key to receiving contributions.
Licensing, code of conduct, governance, and the review and publication process
must all be explicit rather than implicit
to lower the social barriers to contribution.

Tools can help,
especially if they allow proposed changes to be viewed and discussed
prior to their incorporation into the lessons.
(In software development this is known as ``pre-merge review''.)
However,
some tools that are popular in open source software development have considerable up-front learning costs.
Portals like GitHub,
for example,
support everything that open lesson development needs,
but require contributors to use Git,
which has a notoriously steep learning curve \cite{git-survey}.

Complicating matters further,
some file formats make collaboration easier or more difficult.
Despite their ubiquity,
open source version control systems do not directly support review or merge
of Microsoft Office or LibreOffice file formats,
which raises an additional burden for newcomers \cite{jacobs}.
Programmers may look down on Google Docs and wikis
for their lack of pre-merge review and other capabilities,
but their low barrier to entry make them more welcoming to newcomers.

The best way to choose tools for managing lessons is
to ask potential contributors what they are comfortable with
rather than requiring \emph{them} to come to \emph{you}.
Remember also that contributing to a lesson is probably not their top priority,
and look for ways to reduce their cognitive load.
For example,
threaded discussion forums can improve the signal-to-noise ratio
by reducing long reply-all email exchanges.
Several open frameworks are available to facilitate development of new lessons,
such as learnr (\texttt{https://rstudio.github.io/learnr}),
Morea (\texttt{https://morea-framework.github.io}),
and DataCamp's templates (\texttt{https://www.datacamp.com/teach/documentation}).

\rulemajor{Build community around lessons}{community}

Software versions and dependencies are constantly changing,
while the academic literature is advancing at an ever-increasing pace.
As a result,
what is cutting edge one year may be out of date the next and simply wrong the year after.
Collaborative lesson development groups must therefore focus
on creating a community in which contributors support each other
rather than on relying on a small group of stewards.
Authors cannot be expected to maintain continual vigilance on a lesson,
but this is necessary for continual use.

A key part of doing this is to create opportunities for legitimate peripheral participation.
Curating a list of small tasks that newcomers can easily tackle,
encouraging them to give feedback on proposed changes,
or asking them to add new exercises and tweak diagrams and references
can all provide an on-ramp for people who might question their authority or ability to change the main body of a lesson.
Equally,
acknowledging all contributions,
however small,
gives new contributors an early reward for taking part.

Finally,
working in the open can be great,
but can also unintentionally suppress voices.
Programming Historian makes an ombudsperson available for private chats and facilitation
to ensure that no one is excluded.

\rulemajor{Publish periodically and recognize contributions}{publish}

Like software,
specific versions of lessons should be published or released periodically
so that learners or instructors have something stable to refer to for the duration of their use.
Periodic releases also provide an opportunity
for recognizing the contributions of new authors and maintainers.

Academia has only a few ways of recognizing contributions.
Until these are expanded,
it is important to publish lessons in ways that traditional academic systems can digest.
One is to give releases DOIs supplied by providers such as Zenodo (\texttt{https://zenodo.org/})
or DataCite (\texttt{https://www.datacite.org/}).
Contributors can be listed as authors
and the maintainers of the lesson as editors
to differentiate recognition of their contributions.
Each time the lesson is published,
names (and identifiers such as ORCIDs (\texttt{https://orcid.org}))
should be gathered for all contributors.

A lesson release is a good opportunity to bring the material into a stable shape
by fixing outstanding issues and merging contributions.
Version control automatically maintains a list of contributors,
and can also be used to track what content is in what release
(e.g., using branches or tags).
Lesson releases should use a consistent naming scheme;
Software Carpentry has used the year and month of release
(e.g., ``2017.05'')
in its releases \cite{shell2015,shell2017}.

If lessons are being released regularly,
automate the process
and archive old versions in a discoverable location.
Also make sure that everyone involved knows what ``done'' look like,
i.e.,
which outstanding issues have to be addressed
and how it has to be formatted
in order for the next release to go out.
A simple checklist stored with the lesson materials is good enough to start,
but as time goes by,
the community may want to use an issue tracking system of some sort
so that work items can be assigned to specific people
and then ticked off as they are completed.

\rulemajor{Evaluate lessons at several scales}{evaluate}

What people immersed in developing lessons think needs fixing
can easily differ from what learners think.
It is therefore critical to gather and act on feedback at several scales
to check assumptions and stay on course.

Micro-scale feedback can be gathered by an instructor while teaching a particular lesson.
Learners can provide feedback on everything from typographical errors
and the clarity of quiz questions
to the order in which topics are presented,
all of which the instructor should record at the end of each class
in some shared location (such as a Google Doc or GitHub issues).
As well as encouraging direct verbal feedback,
it's a good idea to provide learners with a means to provide feedback anonymously during class
(e.g., on small pieces of paper like sticky notes or through anonymous surveys).

Pre- and post-class surveys and interviews should be used to uncover larger issues,
particularly those arising from developers not fully understanding their audience,
e.g.,
assuming prior knowledge that learners do not have.
Post-class surveys are most effective when conducted 30--90 days after class;
this gives people time to reflect,
so their feedback will more accurately reflect what they learned
rather than how entertained they were.
Clearly-stated learning objectives (\ruleref{practices}) are essential here,
as they tell assessors what they should be measuring.

\rulemajor{Reduce, re-use, recycle}{rrr}

Just as a scholar would not write a paper without a literature review,
an instructor should not create a new lesson if there is an existing one they could use or contribute to.
A short online search can discover if someone has written what you need,
whether it is complementary to your goals,
and if it can be tweaked or modified to meet your needs.

Before re-using content,
make sure to check its license.
Both Programming Historian and the Carpentry projects
use the Creative Commons--Attribution license
(\texttt{https://creativecommons.org/licenses/by/4.0/}),
which allows people to share and adapt material for any purpose
as long as they cite the original source.
Other Creative Commons licenses may restrict commercial use
and/or creation of derivative materials.

The question of licensing also arises when recycling lesson components
such as images, data, figure, or code.
If the license does not cover them explicitly,
ask permission as you would for any other academic material.

The converse of this rule is to make the license for your lessons explicit and discoverable.
For example,
when lessons are published (\ruleref{publish}),
make sure that that keywords such as ``CC-BY'' appear in their bibliography entries
and HTML page headers.

\rulemajor{Link to other resources}{link}

Learners are unlikely to absorb everything they need to know about a topic from your lesson alone.
This is partly a matter of scope---any interesting subject is too large
to fit in a single lesson---but also a matter of level and direction.
As Caulfield has argued \cite{choral-explanations},
the best way to use the Internet is to provide a chorus of explanations
that offer many angles and approaches for any given topic,
each of which may be the best fit for a different set of needs.

Collaboratively-developed lessons should direct learners to these resources at strategic points.
Textbooks,
technical documentation,
videos,
web pages,
threads on Quora or mailing lists:
if a community or discussion forum exists for the topic,
it is worth including.

Doing this is substantial work,
and maintaining it is even more so,
which makes building community around lessons (\ruleref{community}) all the more important.
In particular,
it is vital to engage the learners as equal participants in that community:
they should both be able to propose updates, corrections, and additions to lessons,
and know that they are encouraged to do so (\ruleref{empower}).

\rulemajor{You can't please everyone}{everyone}

No single lesson can be right for every learner.
Two people with no prior knowledge of a specific subject
may still be able to move at different speeds
because of different levels of general background knowledge.
Similarly,
lessons on ecology for learners in Utah and Vietnam
will probably be more relatable if they use different examples.
A community may therefore maintain several differently-oriented or differently-paced lessons
on a single topic,
just as programming languages provide several different libraries for doing the same general thing
with different levels of performance and complexity.

Similarly,
no lesson development community can serve all purposes.
Some groups may prioritize rapid evolution,
while others may prefer a ``measure twice, cut once'' approach.
If there are complementary ways to explain something,
or points of view that can cohabit respectfully,
it may be possible to present them side by side.
There are good pedagogical reasons to do this
even if contributors \emph{do not} disagree:
weighing alternatives fosters higher-order thinking.

But sometimes choices must be made.
The open source software community has wrestled with these issues for three decades,
and has evolved some best practices to address them
\cite{producing-oss}.
As discussed in \ruleref{empower},
the first step is to have a clear governance structure and a clear, permissive license.
Minor disagreements should be discussed openly and respectfully.
If they turn out not to be so minor after all,
contributors should split off and evolve the lesson in the way they see best.
(This is one of the reasons to have a permissive license.)

These splits rarely happen in practice.
When they do,
it is important to remember that we all share the same vision of better lessons, built together.

\section*{Conclusion}

Every day,
teachers all over the world spend countless hours duplicating each other's work.
These ten rules provide an alternative:
adopting the model of collaborative software development
to make more robust and sustainable lessons
that can be continually improved by those who use them.
We hope that our experiences can help others teach more
with more impact and less effort.

% \nolinenumbers

% Either type in your references using
% \begin{thebibliography}{}
% \bibitem{}
% Text
% \end{thebibliography}
%
% or
%
% Compile your BiBTeX database using our plos2015.bst
% style file and paste the contents of your .bbl file
% here. See http://journals.plos.org/plosone/s/latex for
% step-by-step instructions.

\newpage

%%%%% Figure  dissociation
\begin{figure}[ht]  % h = here, t = top, b = bottom, p = float
\includegraphics[width=\linewidth]{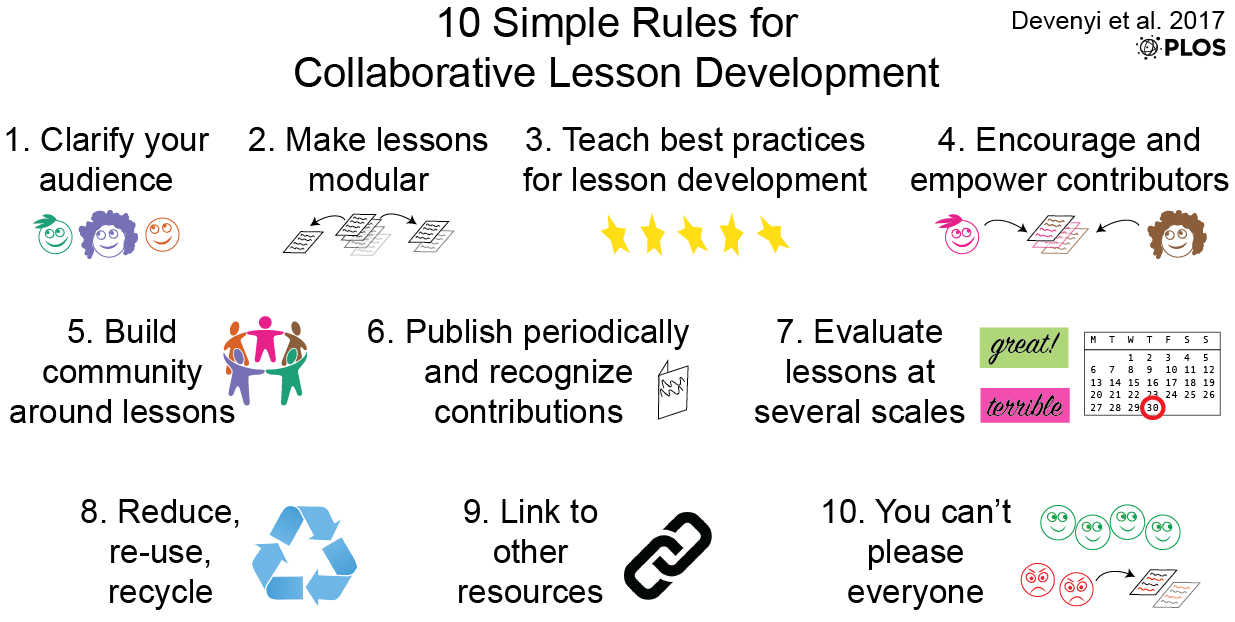}
\caption{Graphical abstract of 10 simple rules for collaborative lesson development}
\label{figure1}
\end{figure}

\newpage

%%%%% Figure  dissociation
\begin{figure}[ht]  % h = here, t = top, b = bottom, p = float
\includegraphics[width=\linewidth]{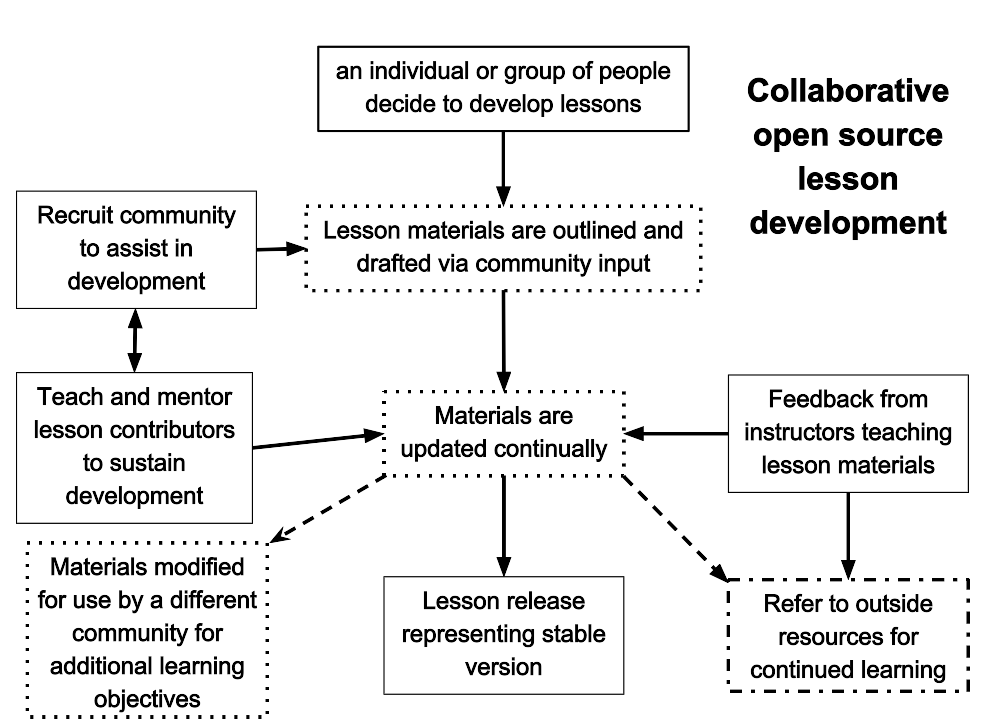}
\caption{Collaborative open lesson development}
\label{figure2}
\end{figure}

\newpage

\begin{mdframed}

\noindent
Box 1: Learner Profile

Jorge has just moved from Costa Rica to Canada to study agricultural engineering.
While fluent in both Spanish and English,
he has a hearing disability that sometimes makes it hard for him to understand lectures,
particularly in noisy environments.
Other than using Excel, Word, and the Internet,
Jorge's most significant previous experience with computers is
helping his sister build a WordPress site for the family business.

Jorge needs to measure properties of soil from nearby farms
using a handheld device that sends text files to his computer.
Right now, Jorge has to open each file in Excel,
crop the first and last points,
and calculate an average.
This workshop will show Jorge how to write a little Python program
to read the data,
select the right values from each file,
and calculate the required statistics.

\end{mdframed}

\end{document}